\documentclass[aps,prl,reprint,showpacs]{revtex4-1}

\usepackage{graphicx}
\usepackage{amsmath}

\usepackage{hyphenat}

\usepackage[]{subcaption} 

\begin{document}


\title{Extreme environmental testing of a rugged correlated photon source}



\author{James A. Grieve}
\author{Robert Bedington}

\author{Alexander Ling}
\email[For correspondance: ]{cqtalej@nus.edu.sg}
\affiliation{Centre for Quantum Technologies, National University of Singapore\\Block S15, 3 Science Drive 2, 117543 Singapore}



\date{April 1st 2015}

\begin{abstract}
Experiments in long distance quantum key distribution have motivated the development of ruggedised single photon sources, capable of producing useful correlations even when removed from the warm, nurturing environment found in most optics laboratories. As part of an ongoing programme to place such devices into low earth orbit (LEO), we have developed and built a number of rugged single photon sources based on spontaneous parametric downconversion. In order to evaluate device reliability, we have subjected our design to various thermal, mechanical and atmospheric stresses. Our results show that while such a device may tolerate launch into orbit, operation in orbit and casual mishandling by graduate students, it is probably unable to survive the forcible disassembly of a launch vehicle at the top of a ball of rapidly expanding and oxidising kerosene and liquid oxygen.
\end{abstract}



\maketitle


\section{Introduction \label{sec:intro}}

Space hardware is required to be tough. In orbit a satellite will experience continual thermal cycling from the unmitigated heat of the sun with no convective cooling available, contrasted with the frozen inky blackness of space when in the shadow of the Earth. Additionally it  will find itself blasted by UV radiation and energetic particles and at the mercy of micrometeorites travelling at the order of 10s km/s.


Perhaps the greatest challenge for the toughness of space hardware however is its initial journey into space and the principal concern here is vibration. Vibrations arise from the process of burning rocket propellant as well as from supersonic effects. Given that most rockets are developed from missiles (where speed of `delivery' is a greater concern than comfort) these vibrations can be considerable. 

Since in-orbit repairs and maintenance in the post-shuttle era are not practical, it is important that the design is right first time. When the space hardware in question is a highly sensitive optics experiment (such as a correlated photon source) a sufficiently ruggedised design is a challenging problem to solve and may require many iterations of Earth-based testing.  

Attractive features for Earth-based testing include the abilities to vary multiple harsh conditions at once in carefully controlled, highly monitored enviroments e.g. performing thermal and vacuum tests simultaneously in thermal-vacuum chambers, or the ability to vibrate along many axes simultaneously. Magnitude is also of utmost importance; the greater the extremes a payload can be designed and tested to, the greater certainty can be had of its successful and long-lived deployment in orbit. 


This short article describes a new testing method that was fortuitously awarded to the authors allowing for potentially much more rigorous qualification testing approach and the development of devices that could almost literally be described as bombproof. \mbox{Such is the serendipity of science.}

\section{Experimental methods \label{sec:methods}}

    A rigorous qualification campaign has proceeded alongside the development of this single photon source. In addition to testing various components individually for their tolerance to radiation~\cite{Radiation15}, we have performed a variety of environmental and thermal tests on the system as a whole.
    
    In 2014, a field test to 35\,km using a high altitude weather balloon was used to test thermal, vacuum and vibration resilience simultaneously, in a space-like environment~\cite{NearSpace14}.
    
       \begin{figure}[h!]
        \includegraphics[width=8cm]{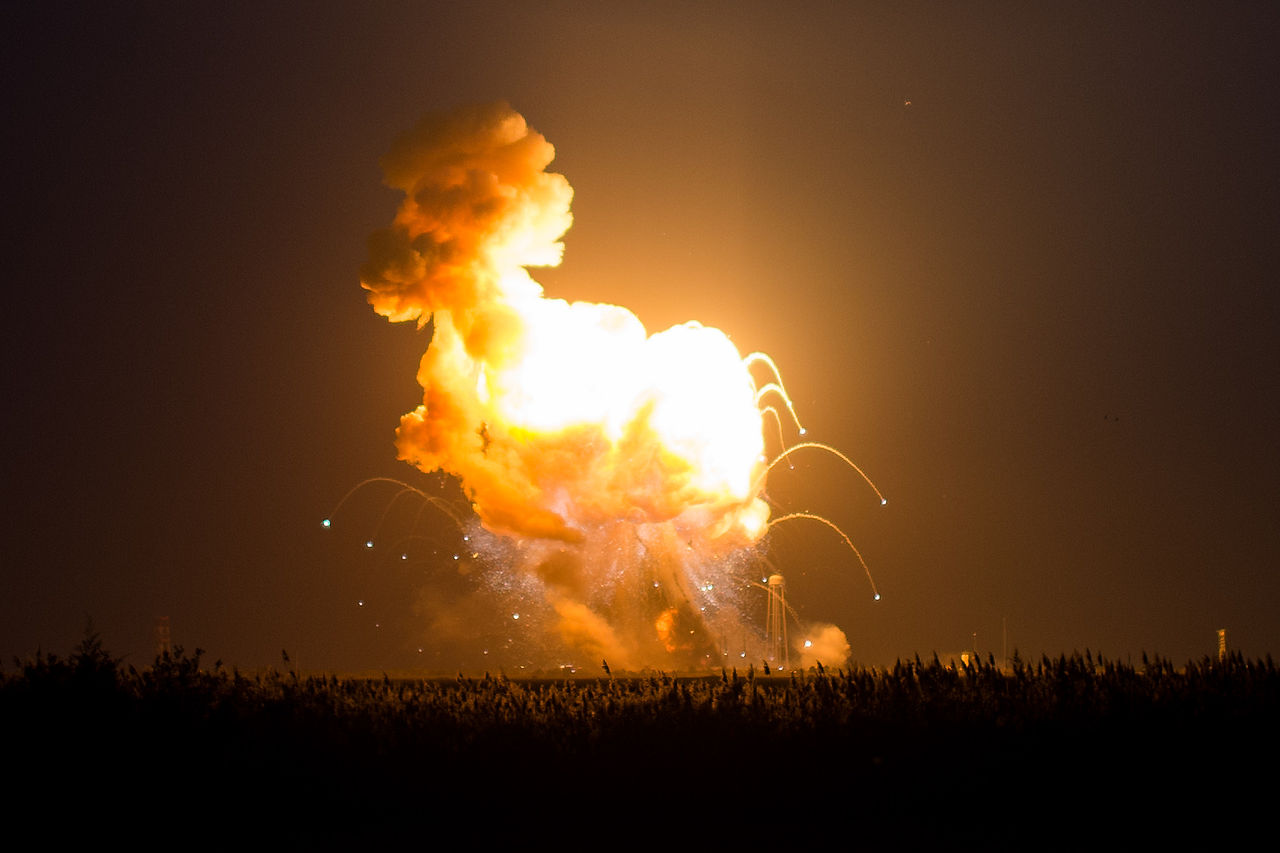}
        \caption{\label{fig:orb3}Fireball associated with the failed Orb3 launch Image courtesy of NASA/Joel Kowsky.}
   \end{figure}
   
   \subsection{Rapid diassembly of a launch vehicle}
   
   In October 2014, the opportunity unexpectedly arose to test our device in the most extreme environment possible: the rapid and unexpected disassembly of a launch vehicle atop a large fireball of kerosene and liquid oxygen. The SPEQS unit was integrated into a nanosatellite (GOMX-2, a 2U CubeSat developed by GOMSpace) and placed into the payload bay of the Cygnus CRS Orb-3 ISS resupply mission. Approximately 15 seconds after ignition, the first stage (Antares 130, Orbital Sciences) suffered a loss of propulsion, resulting in the activation of the flight termination system and the immediate and spectacular recombination of the propellant and liquid oxygen.
   
   It is difficult to calculate the precise temperatures and G-forces this event imparted to the SPEQS payload, but we can attempt to be quantitative nonetheless. The temperature of combustion of the Antares 130's RP-1/LOX mixture is given as 3,670\,K.
   
   Although the temperature and energy released by fireballs have been the subject of previous investigations, the precise accelerations experienced by our payload are hard to calculate. Calculations would necessarily rely on the trajectory the payload took once separated from the main launch vehicle, as well as the mass of other dislodged sections of the cargo it may have been attached to.

\section{Results \& Discussion \label{sec:results}}

\begin{figure}[t!]
        \vspace{1cm}
        \begin{subfigure}{8cm}
            \includegraphics[width=7cm]{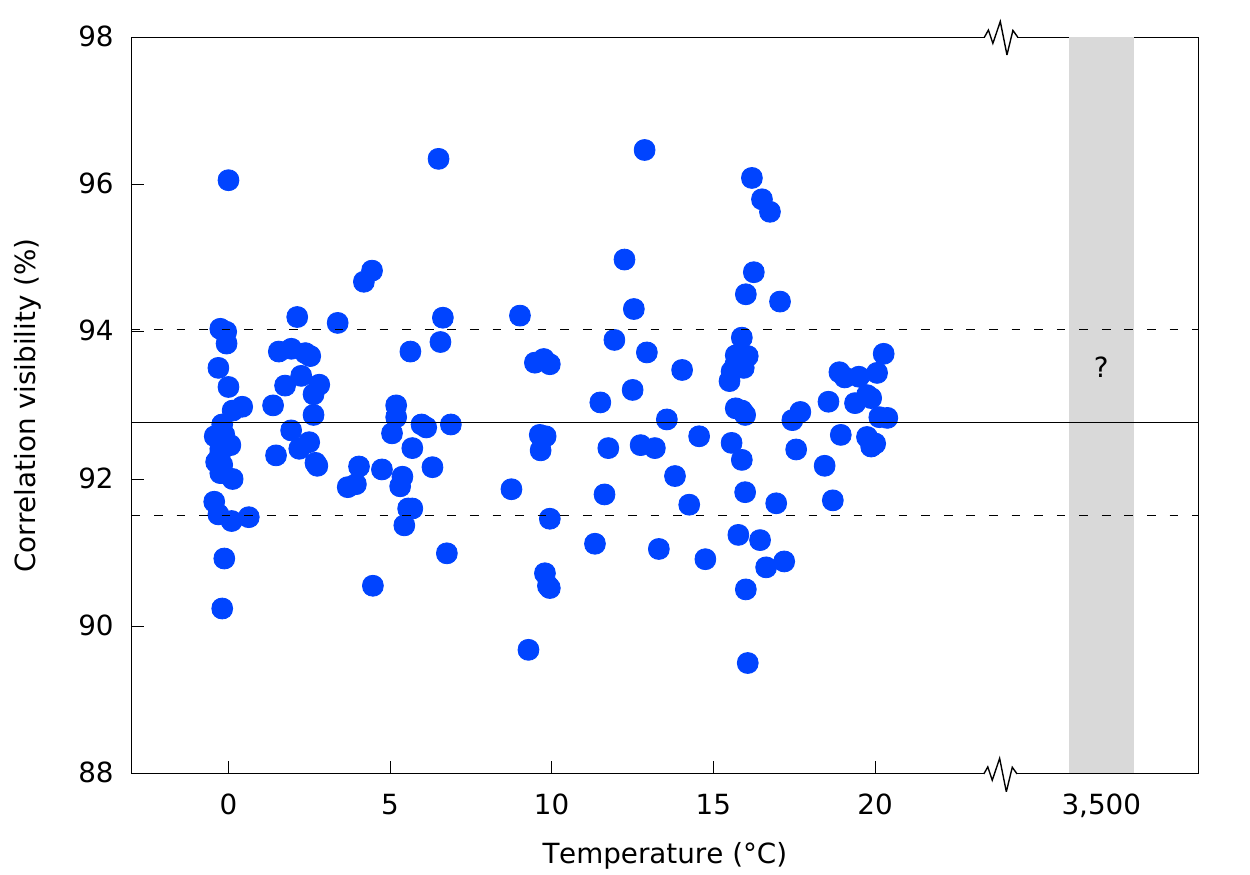}
            \caption{}
        \end{subfigure}
        
        \begin{subfigure}{8cm}
            \includegraphics[width=7cm]{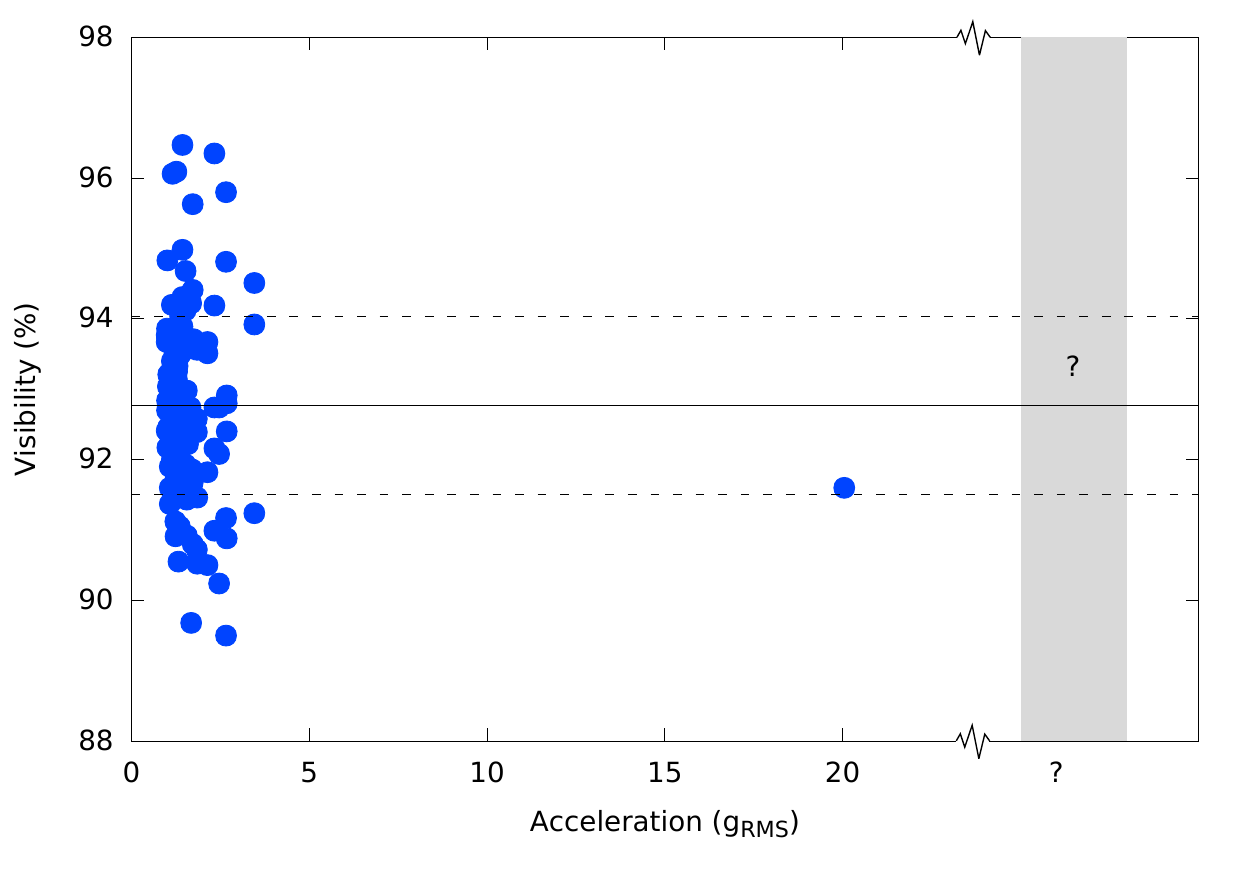}
            \caption{}
        \end{subfigure}
        \caption{\label{fig:vis-temp}Polarization correlation visibility plotted against (a) temperature and (b) acceleration for a similar SPEQS payload. Data is taken from \cite{NearSpace14}, a field test involving a high altitude weather balloon. Solid line indicates the mean visibility, while dashed lines indicate one standard deviation. The grey regions show the approximate temperature and acceleration the payload will have experienced during the Orb-3 explosion, for which unfortunately we have no correlation data.}
    \end{figure}

    Figure~\ref{fig:vis-temp} shows the visibility of our correlated photon source measured at various temperatures and accelerations, with data taken from a high altitude field test~\cite{NearSpace14}. The axes have been expanded to include the probable environment that the payload may have experienced in this more recent environmental testing, though these are necessarily somewhat vague.
    
    Unfortunately, launch regulations prevent the payload from being powered on during the ascent into orbit, and as such we were unable to collect live correlation data during the explosion. The intense heat and difficulty in retrieving the payload also make further operation of the payload impractical, so at this time we do not have any further data to report.

\section{Conclusions \label{sec:conclusion}}

    At the present stage of design maturity, we believe our environmental tests show the SPEQS payload is sufficiently robust to survive launch into low earth orbit, and to perform for its expected lifetime in the space environment (see referneces \cite{NearSpace14, Radiation15}). Although we were able to expose the system to the more demanding environment of a massive kerosene fireball, we were unable to retrieve any data on its performance. For this extremely demanding test to become a routine component of the satellite development toolkit, the data retrieval problem must be overcome. \emph{It is hoped that future launches will not provide opportunities to investigate this further.}

\begin{acknowledgments}
	
The authors would like to thank Zhongkan Tang, Rakhitha Chandrasekara, Yau Yong Sean, Cliff Cheng, Tan Yue Chuan and Christophe Wildfeuer for supplying the weather balloon data.
	
\end{acknowledgments}

\bibliography{april}

\begin{thebibliography}{2}%
\makeatletter
\providecommand \@ifxundefined [1]{%
 \@ifx{#1\undefined}
}%
\providecommand \@ifnum [1]{%
 \ifnum #1\expandafter \@firstoftwo
 \else \expandafter \@secondoftwo
 \fi
}%
\providecommand \@ifx [1]{%
 \ifx #1\expandafter \@firstoftwo
 \else \expandafter \@secondoftwo
 \fi
}%
\providecommand \natexlab [1]{#1}%
\providecommand \enquote  [1]{``#1''}%
\providecommand \bibnamefont  [1]{#1}%
\providecommand \bibfnamefont [1]{#1}%
\providecommand \citenamefont [1]{#1}%
\providecommand \href@noop [0]{\@secondoftwo}%
\providecommand \href [0]{\begingroup \@sanitize@url \@href}%
\providecommand \@href[1]{\@@startlink{#1}\@@href}%
\providecommand \@@href[1]{\endgroup#1\@@endlink}%
\providecommand \@sanitize@url [0]{\catcode `\\12\catcode `\$12\catcode
  `\&12\catcode `\#12\catcode `\^12\catcode `\_12\catcode `\%12\relax}%
\providecommand \@@startlink[1]{}%
\providecommand \@@endlink[0]{}%
\providecommand \url  [0]{\begingroup\@sanitize@url \@url }%
\providecommand \@url [1]{\endgroup\@href {#1}{\urlprefix }}%
\providecommand \urlprefix  [0]{URL }%
\providecommand \Eprint [0]{\href }%
\providecommand \doibase [0]{http://dx.doi.org/}%
\providecommand \selectlanguage [0]{\@gobble}%
\providecommand \bibinfo  [0]{\@secondoftwo}%
\providecommand \bibfield  [0]{\@secondoftwo}%
\providecommand \translation [1]{[#1]}%
\providecommand \BibitemOpen [0]{}%
\providecommand \bibitemStop [0]{}%
\providecommand \bibitemNoStop [0]{.\EOS\space}%
\providecommand \EOS [0]{\spacefactor3000\relax}%
\providecommand \BibitemShut  [1]{\csname bibitem#1\endcsname}%
\let\auto@bib@innerbib\@empty
\bibitem [{\citenamefont {Tan}\ \emph {et~al.}(2015)\citenamefont {Tan},
  \citenamefont {Chandrasekara}, \citenamefont {Cheng},\ and\ \citenamefont
  {Ling}}]{Radiation15}%
  \BibitemOpen
  \bibfield  {author} {\bibinfo {author} {\bibfnamefont {Y.~C.}\ \bibnamefont
  {Tan}}, \bibinfo {author} {\bibfnamefont {R.}~\bibnamefont {Chandrasekara}},
  \bibinfo {author} {\bibfnamefont {C.}~\bibnamefont {Cheng}}, \ and\ \bibinfo
  {author} {\bibfnamefont {A.}~\bibnamefont {Ling}},\ }\href@noop {} {\bibfield
   {journal} {\bibinfo  {journal} {Submitted to the Journal of Modern Optics}\
  } (\bibinfo {year} {2015})}\BibitemShut {NoStop}%
\bibitem [{\citenamefont {Tang}\ \emph {et~al.}(2014)\citenamefont {Tang},
  \citenamefont {Chandrasekara}, \citenamefont {Yau}, \citenamefont {Cheng},
  \citenamefont {Wildfeuer},\ and\ \citenamefont {Ling}}]{NearSpace14}%
  \BibitemOpen
  \bibfield  {author} {\bibinfo {author} {\bibfnamefont {Z.}~\bibnamefont
  {Tang}}, \bibinfo {author} {\bibfnamefont {R.}~\bibnamefont {Chandrasekara}},
  \bibinfo {author} {\bibfnamefont {Y.~S.}\ \bibnamefont {Yau}}, \bibinfo
  {author} {\bibfnamefont {C.}~\bibnamefont {Cheng}}, \bibinfo {author}
  {\bibfnamefont {C.}~\bibnamefont {Wildfeuer}}, \ and\ \bibinfo {author}
  {\bibfnamefont {A.}~\bibnamefont {Ling}},\ }\href@noop {} {\bibfield
  {journal} {\bibinfo  {journal} {Scientific Reports}\ }\textbf {\bibinfo
  {volume} {4}} (\bibinfo {year} {2014})}\BibitemShut {NoStop}%
\end{thebibliography}%

\end{document}